\documentclass[11pt]{article}
\usepackage{amssymb}
\usepackage{amsmath}
\usepackage{amsthm}
\usepackage{amsfonts}
\usepackage{url}
\usepackage{hyperref}
\usepackage{breakurl}
\begin{document}
\bibliographystyle{plain}
\title{{The Myth of Equidistribution\\ for High-Dimensional Simulation}%
\thanks{Copyright \copyright 2007--2010 R.~P.~Brent \hspace*{\fill} rpb240}
}
\author{~\\[10pt]Richard P.\ Brent\\
	Mathematical Sciences Institute\\
	Australian National University\\
	Canberra, ACT 0200, Australia\\[10pt]
	8 May 2010\\[30pt]
\em Presented at a Workshop on\\ 
\em High Dimensional Approximation\\[10pt]
held at the\\
Australian National University\\
19 February 2007}
\date{~}
\maketitle
\thispagestyle{empty}			
\vspace*{\fill}

\subsection*{Abstract}

A pseudo-random number generator (RNG) might be used to generate $w$-bit
random samples in $d$ dimensions if the number of state bits is at least
$dw$. Some RNGs perform better than others and the concept of
equidistribution has been introduced in the literature in order to rank
different RNGs. 
\smallskip

We define what it means for a RNG to be
$(d,w)$-equidistributed, and then argue that $(d,w)$-equidistribution is not
necessarily a desirable property. 

\vspace*{\fill}\pagebreak

\section{Motivation}

\begin{quotation}
{\em
\noindent
There is no such thing as a random number~-- there are only methods to
produce random numbers, and a strict arithmetic procedure of course is not
such a method.} 
\end{quotation}
\rightline{John von Neumann~\cite[p.~768]{Neumann-work}} 
\bigskip

Suppose we are performing a simulation in $d$ dimensions. For simplicity
let the region of interest be the unit hypercube $H = [0,1)^d$.
\smallskip

For the simulation we may need a sequence $y_0, y_1, \ldots$ of points
uniformly and independently distributed in $H$. A pseudo-random number 
generator gives us a sequence $x_0, x_1, \ldots$ of points in $[0,1)$.
Thus, it is natural to group these points in blocks of $d$, that is
\[y_j = (x_{jd}, x_{jd+1}, \ldots, x_{jd + d - 1})\,.\]
If our pseudo-random number generator is good and $d$ is not too large,
we expect the $y_j$ to behave like uniformly and independently distributed
points in $H$.

\section{Pseudo-random vs quasi-random}

We are considering applications where the (pseudo-)random number generator
should, as far as possible, be indistinguishable from a perfectly random
source.  In some applications, e.g.~Monte Carlo quadrature, it is better to
use {\em quasi-random} numbers which are intended for that application and
give an estimate with smaller variance than we could expect with a
perfectly random source.
\smallskip

For example, when estimating a contour integral of an analytic function, we
might transform the contour to a circle and use equally spaced points on the
circle.
\smallskip

However, when simulating Canberra's future climate and water supply, it
would not be a good idea to assume that exceptionally dry years were equally
spaced!

\section{Goodness of fit}

If we use the $\chi^2$ test to test the hypothesis that a set of data is
a random sample from some distribution, then we typically reject the
hypothesis if the $\chi^2$ statistic is {\em too large}. 
\smallskip

However, we
should equally reject the hypothesis if $\chi^2$ is {\em too small}
(because in this case the fit is {\em too good})~\cite{Metropolis50}.

\section{Linear congruential generators}

In the ``old days'' people often followed Lehmer's suggestion and
used linear congruential random number generators
of the form
\[z_{n+1} = az_n + b \bmod m\,.\]
This gives an integer in $[0,m)$ so needs to be scaled:
\[x_n = z_n/m\,.\]
Typically $m$ is a power of two such as $2^{32}$ or $2^{64}$, or a prime
close to such a power of two.
\smallskip

Unfortunately, all such linear congruential generators perform badly
in high dimensions, as shown in Marsaglia's famous paper {\em Random
numbers fall mainly in the planes}~\cite{Marsaglia68}.  

\section{RANDU}

Some linear congruential generators perform disastrously.
For example, consider the infamous RANDU:
\[z_{n+1} = 65539z_n \bmod 2^{31}\]
(with $z_0$ odd). 
These points satisfy
\[z_{n+2} - 6z_{n+1} + 9z_n = 0 \bmod 2^{31}\]
so in dimension $d = 3$ the resulting points $y_j$ all lie on a small
number of planes, in fact $15$ planes separated by distance
$1/\sqrt{1^2 + 6^2 + 9^2} \approx 0.092$
\smallskip

In general, such behaviour is detected by the {\em spectral
test}~\cite{Knuth2}.
\smallskip

Even the best linear congruential generators perform badly because they
have period at most $m$, so the average distance between points $y_j$
is of order \[\frac{1}{m^{1/d}}\] 
(so the set of points closest to any one $y_j$ has volume of order $1/m$). 

\section{Modern generators}

Nowadays, linear congruential generators are rarely used in high-dimensional
simulations. Instead, generators with much longer periods are used. A
popular class is those given by a linear recurrence over $F_2$.
These take the form
\[u_i = Au_{i-1} \bmod 2\]
\[v_i = Bu_i \bmod 2\]
\[x_i = \sum_{j=1}^w v_{i,j}2^{-j}\]
where $u_i$ is an $n$-bit state vector, $v_i$ is a $w$-bit output vector
which may be regarded as a fixed-point number $x_i$,
and the linear algebra is performed over the field $F_2 = {\rm{GF}}(2)$ of
two elements $\{0, 1\}$.  
Here $A$ is an $n \times n$ matrix
and $B$ is a $w \times n$ matrix (both over $F_2$).
Usually $A$ is sparse (so the matrix-vector 
multiplication can be performed quickly) and often $B$ is a projection.

\section{The period}

Provided the characteristic polynomial of $A$ is primitive over 
$F_2$, and $B \ne 0$, the period of such a generator is 
$2^n-1$. This can be very large, e.g. 
$n = 4096$ for {\em xorgens}~\cite{xorgens} and
$n = 19937$ for the {\em Mersenne Twister}~\cite{Matsumoto98}.  
For details we refer to L'Ecuyer's papers~\cite{Ecuyer04,Panneton05}. 

\section{Equidistribution}

Various definitions of $(d,w)$-equidistribution can be found in the literature.
We follow Panneton and L'Ecuyer~\cite{Panneton05} without attempting to
be too general.
\smallskip

Consider $w$-bit fixed-point numbers. There are $2^w$ such numbers
in $[0,1)$. Each such number can be regarded as representing a small
interval of length $2^{-w}$.
\smallskip

Similarly, in $d$ dimensions, we can consider small hypercubes whose
sides have length $2^{-w}$. Each small hypercube has volume $2^{-dw}$
and there are $2^{dw}$ of them in the unit hypercube $[0,1)^d$.
A small hypercube can be specified by a $d$-dimensional vector of
$w$-bit numbers (a total of $dw$ bits).

\subsection*{Definition}

Consider a random number generator with period $2^n$. (A slight change
in the definition can be made to accomodate generators with period $2^n-1$.)
\smallskip

If the generator is run for a
complete period to generate $2^n$ pseudo-random points in $[0,1)^d$,
we say that the generator is $(d,w)$-equidistributed if the same number
of points fall in each small hypercube.
\smallskip

The condition $n \ge dw$ is necessary.  The number of points in each small
hypercube is $2^{n-dw}$.
\smallskip

RANDU (with $n=29$)  
is {\em not} $(d,w)$-equidistributed for any $d \ge 3$, $w \ge 4$. 
However, most good long-period generators {\em are} $(d,w)$-equidistributed
for $dw \ll n$.

\section{Figures of merit}

The maximum $w$ for which a generator can be $(d,w)$-equidistributed
is $w^*_d = \lfloor n/d \rfloor$.  If a generator is actually 
$(d,w)$-equidistributed for $w \le w_d$ then
\[\delta_d = w^*_d - w_d\]
is sometimes called the ``resolution gap''~\cite{Ecuyer04} and
\[\Delta = \max_{d \le n} \delta_d\]
is taken as a figure-of-merit (small $\Delta$ is desirable).
However, this only makes sense when comparing generators with the same
period. When comparing generators with different periods, it makes
more sense to consider
\[W = \sum_{d \le n} w_d\]
as a figure of merit (a large value is desirable).
An upper bound is $W \le \sum_d w^*_d \sim n\ln n$.

\section{Problems with equidistribution}

A test for randomness should (usually) be passed by a perfectly random
source.  
\smallskip

$(d,w)$-equidistribution applies only to a periodic
sequence: we need to know the period
$N = 2^n$ (or $N = 2^n-1$).
A perfectly random source is not periodic, but we can get a periodic 
sequence by taking the first $N$ elements $(y_0,y_1,\ldots,y_{N-1})$ 
and then repeating them ($y_{i+N} = y_i$).
However, this sequence is unlikely to be $(d,w)$-equi-distributed
unless $d$ and $w$ are very small.
\smallskip

Consider the simplest case $dw = n$. There are $N = 2^n$ small hypercubes
and $N!$ ways in which each of these can be hit by exactly one of
$(y_0,\ldots,y_{N-1})$ out of $N^N$ possibilities.  Thus the probability
of equidistribution is
\[\frac{N!}{N^N} \sim \frac{\sqrt{2\pi N}}{\exp(N)}\,.\]
Recall that $N = 2^n$ is typically very large (for example $2^{4096}$)
so $\exp(N)$ is gigantic.

\subsection*{Independence of ordering}

$(d,w)$-equidistribution is independent of the ordering
of $y_0,\ldots,y_{N-1}$. 
\smallskip

Given a $(d,w)$-equidistributed sequence, we can reorder it in any manner
and the new sequence will still be $(d,w)$-equidistributed.
\smallskip

For example, $y_j = j \bmod 2^n$ gives a $(1,n)$-equidistributed sequence.

\subsection*{A common argument}

It is often argued that, when $n$ is large, we will not use the full
sequence of length $N = 2^n$, but just some initial segment of length $M \ll N$.
If $M \ll \sqrt{N}$ then the initial segment may behave like the
initial segment of a random sequence. However, if this is true, what is
the benefit of $(d,w)$-equidistribution? 

\section{Why consider equidistribution?}

The main argument in favour of considering equidistribution seems to be
that, for several popular classes of pseudo-random number generators, we can
test if the sequence is $(d,w)$-equidistributed without actually generating a
complete cycle of length $N$. 
\smallskip

For generators given by a linear recurrence over
$F_2$, $(d,w)$-equidistribution is equivalent to a certain matrix over $F_2$
having full rank. However, the fact that a property is easily checked does
not mean that it is relevant. We actually need something weaker (but harder
to check).

\section{Conclusion}

When comparing modern long-period pseudo-random number generators,
$(d,w)$-equidistribution is irrelevant, because it is neither necessary
nor sufficient for a good generator.

\bigskip 

\end{document}